\documentclass[12pt]{article}
\usepackage{amsmath}
\usepackage{graphicx,psfrag,epsf}
\usepackage{enumerate}
\usepackage{natbib}
\usepackage{url} 
\usepackage{amsthm}
\usepackage{amsfonts}
\usepackage{bm}
\usepackage{url}
\usepackage{subcaption}
\usepackage{hyperref}
\hypersetup{
    colorlinks,
    citecolor=[rgb]{0,0,.22},
    filecolor=magenta,
    linkcolor=[rgb]{0,.22,0},
    urlcolor=blue
}

\newcommand{\blind}{0}

\begin{document}

\def\spacingset#1{\renewcommand{\baselinestretch}%
{#1}\small\normalsize} \spacingset{1}

\newcommand{\R}{\mathbb R}
\newcommand{\xb}{\boldsymbol{x}}
\newcommand{\yb}{\boldsymbol{y}}
\newcommand{\Xb}{\boldsymbol{X}}
\newcommand{\zb}{\boldsymbol{z}}
\newcommand{\Zb}{\boldsymbol{Z}}
\newcommand{\Pb}{\boldsymbol{P}}
\newcommand{\Lb}{\boldsymbol{L}}

\newcommand{\hbm}{\hat{\boldsymbol{\mu}}}
\newcommand{\hbS}{\hat{\boldsymbol{\Sigma}}}
\newcommand{\bS}{\boldsymbol{\Sigma}}
\newcommand{\ab}{\boldsymbol{a}}
\newcommand{\eb}{\boldsymbol{e}}
\newcommand{\wb}{\boldsymbol{w}}
\newcommand{\hbmr}{\hat{\boldsymbol{\mu}}_\textrm{r}}
\newcommand{\hbSr}{\hat{\boldsymbol{\Sigma}}_\textrm{r}}
\newcommand{\argmin}{\mathop{\mbox{argmin}}}
\newcommand{\thetab}{\boldsymbol{\theta}}
\newcommand{\betab}{\boldsymbol{\beta}}

\theoremstyle{plain}
\newtheorem{thm}{Theorem}[section]
\newtheorem{lem}[thm]{Lemma}
\newtheorem{prop}[thm]{Proposition}
\newtheorem*{cor}{Corollary}

\theoremstyle{definition}
\newtheorem{defn}{Definition}[section]
\newtheorem{exmp}{Example}[section]



\if0\blind
{
  \title{\bf Outlyingness: why do outliers lie out?}
  \author{Michiel Debruyne\\
    Dexia, Belgium\\
    Sebastiaan H\"oppner \\
    Department of Mathematics, KU Leuven\\
    Sven Serneels\\
    BASF Corp.\\
    and\\
    Tim Verdonck\thanks{
    This work was supported by the BNP Paribas Fortis Chair in Fraud Analytics and Internal Funds KU Leuven under Grant C16/15/068. We thank Dries Cornilly for his constructive comments.}\\
    Department of Mathematics, KU Leuven}
  \maketitle
} \fi

\if1\blind
{
  \bigskip
  \bigskip
  \bigskip
  \begin{center}
    {\LARGE\bf Outlyingness: why do outliers lie out?}
\end{center}
  \medskip
} \fi
\bigskip
\begin{abstract}
Outlier detection is an inevitable step to most statistical data analyses. However, the mere {\em detection} of an outlying case does not always answer all scientific questions associated with that data point. Outlier detection techniques, classical and robust alike, will typically flag the entire case as outlying, or attribute a specific case weight to the entire case. In practice, particularly in high dimensional data, the outlier will most likely not be outlying along all of its variables, but just along a subset of them. If so, the scientific question why the case has been flagged as an outlier becomes of interest. In this article, a fast and efficient method is proposed to detect variables that contribute most to an outlier's outlyingness. Thereby, it helps the analyst understand why an outlier lies out. \\
The approach pursued in this work is to estimate the univariate direction of maximal outlyingness. It is shown that the problem of estimating that direction can be rewritten as the normed solution of a classical least squares regression problem. Identifying the subset of variables contributing most to outlyingness, can thus be achieved by estimating the associated least squares problem in a sparse manner. From a practical perspective, sparse partial least squares (SPLS) regression, preferably by the fast sparse NIPALS (SNIPLS) algorithm, is suggested to tackle that problem. The proposed methodology is illustrated to perform well both on simulated data and real life examples. 
\end{abstract}

\noindent%
\vfill

\newpage
\spacingset{1.45} 
\section{Introduction}
\label{sec:intro}

Statistical analysis usually encompasses a step in which outliers need to be processed. What happens to them, depends on the application. Potentially, one is only interested in fitting a model for the bulk of the data, in which case outlier removal fits the purpose, given the outliers have correctly been detected. However, often one would like to know more about these outliers: are they manual errors or measurement errors, or are they just extreme values occurring naturally? Possibly even the outliers belong to separate clusters in the data, previously unassumed? As data dimensions increase, it becomes more likely that outliers of any of these natures will be predominantly outlying only with respect to a subset of the variables they consist of. Ample methodology exists to detect outliers. In this article, methodology will be developed to analyze why outliers lie out, given they have been detected by an appropriate statistic. Consider detection of transfer fraud as an example where the methodology proposed in this article, can have a great practical advantage. Fraud detection is all about outlier detection: typically only few transactions out of a vast number are fraudulent. Therefore, the outliers are the cases of highest interest. Once fraudulent transactions have been detected, one wants to investigate in which way these transactions are suspicious. A method that explains a fraudulent transaction's outlyingness, can speed up that analysis significantly, or even automate it.  

The aim of nonrobust, or {\em classical} statistical methods, such as maximum likelihood or least squares techniques, is to optimally fit an assumed model to all observations in the data. However, real data often contain outliers, i.e. observations that deviate from the assumed model. In their presence, classical methods may become unreliable. Therefore, robust high-breakdown methods have been developed that are not heavily influenced by outliers. These robust alternatives can still reliably estimate the parameters of the postulated model, while a minority (i.e. less than $50\%$) of the data are allowed to deviate arbitrarily far from this model. As an additional benefit, one can detect the outliers as the observations that deviate substantially from the robust fit \citep{Rousseeuw:RobReg}. Note that the outliers are often not detected using the classical fit, since this fit itself is also influenced by these atypical observations, an effect known as {\em masking}. Moreover, the effect of the outliers on a nonrobust fit can be so large that some regular observations may appear to be outlying, which is called {\em swamping} \citep{Davies:Multiple}.

Nowadays, many robust statistical methods are available that are able to detect outliers in multivariate data, both in high and low dimensions \citep{Maronna:RobStat}.
Popular robust mean and covariance estimators are, for example, the MCD estimator \citep{Rousseeuw:LMS,Rousseeuw:FastMCD}, S-estimators \citep{Rousseeuw:RobReg} and $\tau$-estimators \citep{Lopuhaa:TauEst}. When the dimension exceeds the sample size, one can use the OGK estimator \citep{Maronna:OGK} or the MRCD estimator \citep{boudt2017minimum}. Alternatively, a robust PCA method (e.g. \citet{Hubert:ROBPCA,Croux:Proj}) can be applied to detect outliers.

It is important to note that the detected outliers are not necessarily errors in the data. The presence of outliers may reveal that the data is more heterogeneous than previously assumed and also more heterogeneous than what could be handled by the original statistical model. Outliers can be isolated or may come in clusters, indicating that there are subgroups in the population that behave differently. Sometimes outliers can even be the most interesting cases in the entire sample. Robust analysis can provide a better insight in the structure of the data and reveal structures in the data that would remain hidden in a classical analysis. 

The robust estimation methods described above, as well as outlier detection techniques based on classical statistics, typically flag entire cases as outliers. In reality, outliers may only be outlying with respect to a small subset of the variables they consist of. A question that, up to our knowledge, remains unanswered in the robust statistical literature, is the following: once an outlier has been detected in a multivariate data set, how can the subset of variables that contribute most to its outlyingness be identified? Some outliers may be deviating along all of the variables, whereas other outliers may only deviate along just a few of them. Robust statistics treat such outliers in exactly the same way: they down-weight the entire observation. However, if an outlier is only deviating along some variables, it might be more useful to only adjust the atypical values in these variables. In this way, the non-contaminated and potentially valuable information in the other variables is retained. This has become more important in recent years since technical advances have led to the availability of (very) high-dimensional data sets. For instance in genetics, it is perfectly reasonable that an observation deviates from the majority of data points only for a few genes, not for all of them. Obviously, finding this subset of genes would be of high practical interest. Note that in practice, this would imply finding a subset of a few out of several hundreds of thousands of genes.  Similar examples can be found in climatology, geology, neurology, process and analytical chemistry, economics and finance, among others.
This extra information may be very interesting and useful for gaining insight in the data. 
By studying the selected variables, one can explain why a certain outlier is deviating from the pattern of the majority of the observations. 

In order to investigate an outlier's outlyingness, consider the following problem: given a multivariate data set $\boldsymbol{X}=(\xb_1,\ldots, \xb_n)^T$ and the fact that observation $\xb_i\in \R^p$ is an outlier with large outlyingness, find the subset of variables contributing most to the outlyingness of $\xb_i$. This problem is akin to variable selection, with the objective of determining those variables contributing most to outlyingness instead of to predictive power. A simple idea to find relevant variables is to check the univariate direction in which the observation is most outlying. In Section \ref{sec:outregprob}, it is shown that the problem of estimating this direction of maximal outlyingness can be rewritten as the normed solution of a classical least squares regression problem. The proofs of the propositions found there, are given in Appendix \ref{sec:AppA}. Thanks to this result, identifying the subset of variables that contribute most to outlyingness becomes a variable selection problem: investigating the direction of maximal outlyingness is equivalent to investigating the vector regression coefficients of the associated regression problem. Therefore, any widely accepted method for variable selection can be applied to this associated regression problem, ranging from visual inspection of normalized regression coefficients to application of sparse estimation procedures. The latter may be preferable in automated analyses, yet one should keep in mind that these methods still depend on the selection of a sparsity parameter. Various methods exist that allow to estimate the vector of regression coefficients in a sparse way: the lasso \citep{tibshirani1996regression} or the elastic net \citep{zou2005regularization}, for instance, would be suitable to accomplish this task. In this article, however, it is suggested to apply sparse partial least squares (SPLS) regression \citep{chun2010sparse} for these purposes. It has the advantage that a single model component suffices to detect the relevant variables specifically for the outlyingness problem, which is illustrated in the simulation study reported in Section \ref{sec:simstudy}. Moreover, thanks to the univariate nature of the regression problem, SPLS can be calculated by the sparse NIPALS (SNIPLS) algorithm (as first described by \cite{hoffmann2016sparse}), which has the advantage of using exact PLS solutions instead of the numerical optimization applied in \citep{chun2010sparse}. Both of these advantages make the SNIPLS algorithm up to our knowledge the computationally most elegant and efficient way to search for variables contributing to outlyingness in individual cases. Computational efficiency is an important positive property, since this method will realistically be applied to every single outlying case of the data.

The article is organized as follows. In Section \ref{sec:outregprob}, the direction of maximal outlyingness is defined, and its alternative formulation as a least squares regression problem, is introduced.  Section \ref{sec:sparsePLS} outlines how to search for variables contributing to outlyingness by estimating the regression coefficients of the corresponding regression problem through SNIPLS. Section \ref{sec:opteta} describes several graphical tools as well as an automatic approach for selecting the optimal value for the sparsity parameter. In Section \ref{sec:simstudy}, the validity of the approach is illustrated in an extensive simulation study. In Section \ref{sec:example}, the method is applied to real life data. Finally, Section \ref{sec:conclusion} concludes.  

\section{Outlyingness as a regression problem \label{sec:outregprob}}

Let $\Xb=(\xb_1,\ldots,\xb_n)^T$ be an $n\times p$ data matrix with $\xb_i=(x_{i1},\ldots,x_{ip})^T$ and $X_j=(x_{1j},\ldots,x_{nj})^T$ respectively the $i$th row and the $j$th column of $\Xb$ (note that both are column vectors). Denote by $\hbm_r$ and $\hbS_r$ robust estimates of location and scatter for $\Xb$. 
One can then compute the squared robust Mahalanobis distance for every point $\xb\in\mathbb{R}^p$ as
\begin{equation}\label{robustsquaredMahalanobis}
m(\xb;\hbm_r,\hbS_r)^2 = \left(\xb - \hbm_r\right)^T \hbS_r^{-1} \left(\xb - \hbm_r\right).
\end{equation}
These robust Mahalanobis distances measure the distance between point $\xb$ and the robust center taking into account the covariance structure of the data. 

For fat data (i.e. $p>n$), the sample covariance matrix is not existing (since some of its eigenvalues will be zero) and many robust alternatives (such as the MCD estimator) can not be computed. Note that the OGK estimator \citep{Maronna:OGK}, the MRCD estimator \citep{boudt2017minimum} and the robust precision matrix (i.e. inverse of the scatter matrix) of \cite{ollerer2015robust} can still be used. 
As an alternative, one can also use a robust principal component analysis (PCA) method (e.g. \citet{Hubert:ROBPCA,Croux:Proj}),  to obtain a spectral decomposition of the covariance matrix as $\Pb\Lb\Pb^T$ where $\Pb$ and $\Lb$ respectively contain the eigenvectors and eigenvalues of the covariance matrix.  

Based on these distances, a weight $w_i$ can then be assigned to each observation $\xb_i$, indicating whether the observation is outlying or not. Under the assumption of multivariate normal data, squared Mahalanobis distances are asymptotically $\chi^2_p$ distributed \citep{bibby1979multivariate}. Therefore, weights for each observation are often obtained as follows:
\begin{equation} \label{weights}
w_i = \left\{ \begin{matrix} 1  & \mbox{if }m(\xb_i;\hbm_r,\hbS_r)^2 \leq \chi_{p,0.975}^2 \\ 0 & \mbox{otherwise} \end{matrix}\right. .
\end{equation}
Note that other weight functions can of course be used and let $\sum_{i=1}^n w_i=n_w$.
These weights can then be used to compute a weighted mean and weighted covariance matrix:
\begin{align} \label{weightedestimators}
\hbm_w &= \frac{1}{n_w}\sum_{i=1}^{n} w_i\xb_i, \\
\hbS_w &= \frac{1}{n_w-1}\sum_{i=1}^n w_i(\xb_i - \hbm_w)(\xb_i -\hbm_w )^T. \label{CovWeightedMCD}
\end{align}
The outlyingness of a point $\xb\in\mathbb{R}^p$ is defined as the robust Mahalanobis distance using the weighted mean and weighted covariance matrix:
\begin{equation}\label{outlyingness}
o(\xb)^2 = m\left(\xb; \hbm_w, \hbS_w\right)^2 = \left(\xb-\hbm_w\right)^T \hbS_w^{-1} \left(\xb-\hbm_w\right).
\end{equation}
The following proposition is well known in the unweighted case. In Appendix \ref{sec:AppA1}, it is proven that it also holds in the weighted case.
\begin{prop}\label{prop1}
The outlyingness of any point $\xb\in\mathbb{R}^p$ can be expressed as the solution of a maximization problem as follows:
\begin{equation} \label{outlymax}
o(\xb) = \max_{\ab \in \R^p, \|\ab\|=1} \frac{|\xb^T\ab - \hbm_w^T\ab|}{\sqrt{\ab^T\hbS_w\ab}}.
\end{equation}
and the direction $\ab$ that maximizes the right-hand side of the above equation, is equal to 
$$\ab = \frac{\hbS_w^{-1}(\xb - \hbm_w)}{\|\hbS_w^{-1}(\xb - \hbm_w)\|}.$$
\end{prop}

This can be interpreted as searching for the direction $\ab$ such that the distance between the projected point $\xb^T\ab$ and the projected weighted mean $\hbm_w^T\ab$, standardized by a measure of spread of the projected observations, is maximal. The direction for which the maximum in proposition \ref{prop1} is attained, is the {\em direction of maximal outlyingness} for point $\xb$ and will be denoted by $\ab(\xb)$. This direction of maximal outlyingness is potentially interesting, because its coefficients reflect how individual variables contribute to the outlyingness of a point. 

The direction of maximal outlyingness can alternatively be expressed as a normalized least squares problem.

\begin{thm}
	\label{theorem1}
	Let $\xb$ be an arbitrary point in $\mathbb{R}^p$ and $\varepsilon\in\mathbb{R}$ with $\varepsilon>0$. Denote $\boldsymbol{y}_{w,\varepsilon} = \eb_{n+1}$ with $\eb_{n+1}$ the $(n+1)$th basis vector in $\R^{n+1}$ containing $1$ at component $(n+1)$ and $0$ elsewhere. Let $n_{w,\varepsilon}=n_w+\varepsilon$ and $$\hat{\boldsymbol{\mu}}_{w,\varepsilon}=\frac{1}{n_{w,\varepsilon}}\left(\sum_{i=1}^{n} w_i\xb_i + \varepsilon\xb\right).$$ Let $\boldsymbol{X}_{w,\varepsilon} = (\sqrt{w_1}(\boldsymbol{x}_{1} - \hat{\boldsymbol{\mu}}_{w,\varepsilon})^T, \ldots , \sqrt{w_n}(\boldsymbol{x}_{n} - \hat{\boldsymbol{\mu}}_{w,\varepsilon})^T, \sqrt{\varepsilon}(\xb-\hat{\boldsymbol{\mu}}_{w,\varepsilon})^T)^T$, the weighted data to which the row $\sqrt{\varepsilon}(\xb-\hat{\boldsymbol{\mu}}_{w,\varepsilon})^T$ is added, centred around the robust location estimate. Then
	\begin{equation}\label{eq:outregnonsparse}
	\ab(\xb) = \lim_{\varepsilon\rightarrow0} \frac{\thetab_{\varepsilon}}{\|\thetab_{\varepsilon}\|}, \text{ with } \thetab_{\varepsilon} = \argmin_{\betab \in \R^p} \|\boldsymbol{y}_{w,\varepsilon} - \boldsymbol{X}_{w,\varepsilon}\betab\|^2.
	\end{equation} 
\end{thm}
\vspace{4mm}
For proof of Theorem \ref{theorem1}, the reader is referred to Appendix \ref{sec:AppA2}. 
Note that the definition of $\ab(\xb)$ as a limit for $\varepsilon\rightarrow0$ also holds when $\xb\in\{\xb_1,\ldots,\xb_n\}$, in which case
$$\ab(\xb_i) = \frac{\thetab}{\|\thetab\|}, \text{ with } \thetab = \argmin_{\betab \in \R^p} \|\boldsymbol{y}_w - \boldsymbol{X}_w\betab\|^2.$$
where $\yb_w$ is the $i$th basis vector in $\R^n$ and $\boldsymbol{X}_w = (\sqrt{w_1}(\boldsymbol{x}_{1} - \hat{\boldsymbol{\mu}}_{w})^T, \ldots , \sqrt{w_n}(\boldsymbol{x}_{n} - \hat{\boldsymbol{\mu}}_{w})^T)^T$. 
Many robust estimators will assign exact zero case weights to outliers far away from the data centre. Note that if our case of interest is assigned a zero weight ($w_i=0$), then this can be circumvented by replacing the zero weight by a very small weight (e.g. 0.0001). This is equivalent with adding observation $i$ to the data matrix $\Xb$ and assigning it a small weight $\varepsilon$.

Owing to Theorem \ref{theorem1}, outlyingness can be estimated by calculating the vector of regression coefficients $\betab$ in this problem. Different regression estimators can now be plugged in to estimate $\betab$, and as such, outlyingness. The most straightforward choice for a plug in regression estimate is least squares regression. Interpreting which variables contribute most to outlyingness, can then be done by examining the absolute magnitude of the standardized least squares regression coefficients. In practice, however, this can be a tedious process and is challenging to incorporate in an automated procedure.  

Moreover, least squares regression has several important drawbacks. At first, when the number of variables exceeds the sample size, the least squares fit is not well defined and cannot be calculated. Another problem frequently encountered in practice is multicollinearity. Even when some regressors are nearly collinear, it is well known that the results obtained from least squares become unstable. 
Moreover, least squares regression is not sparse, which implies that it typically yields a set of regression coefficients with very few non-zero elements, or none at all. As dimensions increase, this complicates interpretation and is challenging to automate. How to go about these issues, will be discussed in the next Section.

\section{Sparse direction of maximal outlyingness\label{sec:sparsePLS}}

In order to obtain an estimate of the direction of maximal outlyingness that can (i) easily be interpreted and (ii) from which automatically the non-zero elements can be selected, a regression plug-in estimate should be applied to Equation \eqref{eq:outregnonsparse}, that has the capability to produce a sparse vector of regression coefficients. Plenty sparse regression estimators have been described in the literature. These estimators all have in common that they can yield sparse regression coefficients by including a term in their respective objective functions that puts a penalty on the norm of these regression coefficients. The idea of such a penalization goes back to ridge regression \citep{hoerl1970ridge}, where an $L_2$-penalty term on the Euclidean norm of the parameter vector is imposed. This effectively solves ill-posed problems in least squares regression, such as the ones discussed at the end of the previous Section. 

Applying ridge regression to estimate the vector of regression coefficients $\betab$ in \eqref{eq:outregnonsparse}, actually yields an entire path of regularized directions of maximal outlyingness as follows:
\begin{defn}
A path of regularized directions of maximal outlyingness $\ab(\lambda,\xb_i)$ is defined by
\begin{equation} \label{pathregularizeddirection}
\ab(\lambda,\xb_i) = \frac{\thetab(\lambda)}{\|\thetab(\lambda)\|}, \text{ with } \thetab(\lambda) = \argmin_{\betab \in \R^p} \left\{ \|\boldsymbol{y}_w - \boldsymbol{X}_w\betab\|^2 + \lambda \sum_{j=1}^{p} \betab_i^2 \right\},
\end{equation}
\end{defn}
Once the path $\boldsymbol{a}(\lambda, \xb_i)$ is obtained, the objective is to select
a subset of $k$ variables contributing most to the outlyingness. 

Ridge regression, however, does not yield a set of regression coefficients with a subset of elements exactly equal to zero.
Since it cannot produce parsimonious models, alternative, sparse plug-in regression estimators have to be considered. \citet{tibshirani1996regression} has proposed the LASSO which uses $L_1$-norm regularization to effectively shrink many parameter estimates to zero and hence perform an intrinsic variable selection. Other penalty methods that yield sparse models can be applied as well, e.g. the SCAD penalty \citep{fan2001variable}, the minimax concave penalty \citep{zhang2010nearly}, the adaptive lasso \citep{zou2006adaptive} or the Dantzig selector \citep{candes2007dantzig}. The elastic net \citep{zou2005regularization} combines the lasso and ridge penalties to obtain a method that can provide sparse model estimates in the presence of multicollinearity. Among these methods, the LASSO is one of the most frequently applied techniques. 

Using the LASSO as a plug-in estimate into Equation \eqref{eq:outregnonsparse}, actually corresponds to a path of sparse (and still regularized) directions of maximal outlyingness: 
\begin{defn}
A path of sparse directions of maximal outlyingness $\ab(\lambda,\xb_i)$ is defined by
\begin{equation} \label{pathsparsedirection}
\ab(\lambda,\xb_i) = \frac{\thetab(\lambda)}{\|\thetab(\lambda)\|}, \text{ with } \thetab(\lambda) = \argmin_{\betab \in \R^p} \left\{ \|\boldsymbol{y}_w - \boldsymbol{X}_w\betab\|^2 + \lambda \sum_{j=1}^{p} |\betab_i| \right\}. 
\end{equation}
\end{defn}
This path of sparse directions of maximal outlyingness tackles the issue with interpretability of the direction of maximal outlyingness. Yet from a computational perspective, it can still be burthensome. Recall that the procedure should be applied to every single outlying case in a data set. In order to have both the benefits of interpretability (many non-zero elements) and computational elegance, the approach pursued in this work is to combine dimension reduction and a penalty term. This is accomplished by applying sparse partial least squares (SPLS) regression \citep{chun2010sparse} as the plug-in regression estimate into Equation \eqref{eq:outregnonsparse}. Should one plug in SPLS with a maximal number of latent variables, then this approach becomes very similar to scanning a LASSO based path such as defined in Equation \eqref{pathsparsedirection}. Yet the elegance SPLS offers over the other methods, is that it can actually be applied with fewer, or just one, latent variable, without losing interpretative power. Application of SPLS with few latent variables is computationally very efficient and yields good and reliable results for high-dimensional data in practice.

The reason why SPLS performs well in this context, can be interpreted reflecting on how PLS and its sparse counterpart have been conceived.  Partial least squares regression (PLS) is a regression method developed in the 1960s \citep{wold1966estimation} that is particularly suited to model data where the number of variables exceeds the number of cases, as well as multicollinear data. PLS thanks these properties to its implicit dimension reduction step, wherein it typically decomposes the original data $\boldsymbol{X} \in \R^{n \times p}$ onto a subset of $h << p$ latent variables $\boldsymbol{T}$. The latent variables are defined according to a criterion that maximizes covariance with the predictand, which ensures that the latent variables capture a maximal amount of information in the data relevant for prediction. 

The regression problem to estimate outlyingness is particular in the sense that the dependent variable is the unit vector in the $n$ dimensional space, which only has one nonzero entry in the cell that corresponds to the outlying case. Therefore, one can assume that few partial least squares components should be able to capture all variance in the data relevant for predicting this atypical $\boldsymbol{y}$ vector. It is not unreasonable to expect that a single PLS component will always capture a sufficient amount of information for the particular task of outlyingness estimation. This assumption has in fact been corroborated in the course of establishing the results presented in the simulation study (Section \ref{sec:simstudy}).

Partial least squares has the drawback, however, that it is non-sparse, which implies that the vector of regression coefficients will only seldomly have nonzero entries. One could put a threshold on the absolute magnitude of the individual coefficients to determine which variables contribute most to outlyingness. However, sparse partial least squares offers a more elegant alternative, yielding a model consistent estimate for the vector of regression coefficients that is based on a PLS-alike dimensionality reduction on the one hand, but also consistently has a subset of nonzero entries thanks to a sparsity penalty $\eta$ being imposed to the weighting vector (as long as $\eta > 0$). Here, $\eta\in[0,1)$ plays the role of the sparsity parameter and was introduced by \cite{chun2010sparse} to facilitate the parameter selection since the range of $\eta$ is known. 

Sparse partial least squares regression has two drawbacks: on the one hand, its intrinsic minimization may be time consuming, and secondly, it depends on two parameters to be optimized: the sparsity parameter and the number of latent variables. Owing to the univariate nature of the predictand, the former drawback can be avoided by applying the sparse NIPALS (SNIPLS) algorithm instead of the algorithm described in the original paper by \cite{chun2010sparse}. For a univariate predictand, the SNIPLS algorithm is equivalent to the SPLS algorithm, but it is significantly more efficient from a computational perspective. The SNIPLS algorithm was published as an internal subroutine used in the construction of the SPRM-DA classifier \citep{hoffmann2016sparse}, and is also used as an internal step for computing SPRM regression \citep{hoffmann2015sparse}. 

In what follows, it will be described how to select the optimal sparsity parameter. 

\section{Determining the optimal SPLS sparsity parameter\label{sec:opteta}}
The optimal sparsity parameter $\eta$ is the value for which the minimal number of variables is selected, such that the reduced case (i.e. the observation after removing those selected columns from the data set) is no longer outlying (in the lower dimensional feature space).
This optimal SPLS parameter combination has to be determined from the data. For $\eta=0$ the model is estimated including all variables and for $\eta$ close to 1, almost all variables are equal to zero. Therefore, typically a grid of values for $\eta\in[0,1)$ is searched. In Subsection \ref{sec:optimal}, an automatic approach to determine the optimal $\eta$ is described, which leads up to the SPADIMO algorithm (SPArse DIrections of Maximal Outlyingness). In Subsection \ref{sec:graphics}, two graphical tools are presented that give more insight about the selection of the parameter $\eta$. 

\subsection{SPADIMO procedure\label{sec:optimal}}
The approach followed by the authors is the following. From the perspective of the number of variables retained, SPLS converges from none (or some) to all variables in two dimensions: when keeping $\eta$ constant, increasing $h$ will eventually yield a model with nonzero entries for all variables. Likewise, a model with constant $h$ will eventually use all predictors available as $\eta$ approaches zero. Based on the conjecture that an SPLS model with one latent variable should be able to capture all variance in the data relevant to predict a unit vector, it is plausible to fix the SPLS number of latent variables to one, and then screen $\eta$ in a given range from high to low. This order of proceeding will make sure that in the first iteration, the most sparse estimate is constructed, based on none to just a few of the original set of variables. 

It then becomes a good question at which value of $\eta$ to terminate the algorithm. For that purpose, consider the following remark. If the case consisted entirely of cells that fall within the bulk of the data, it would not be flagged as an outlier. Therefore, it makes sense to proceed as follows: for each $\eta$, compute an SPLS regression estimate (with $h$=1) determining a subset of variable(s) contributing to outlyingness. Then, replace the cells in the outlier corresponding to those variables by missing values, and estimate the case's outlyingness. In order to be able to estimate the outlyingness, the missing values need to be removed from the data. A viable way to do this is by omitting the entire columns corresponding to those variables that have already been detected as contributors to outlyingness. After applying a robust estimator for location and scale on this data (or robust PCA) one calculates the weights as in formula (\ref{weights}) and then obtain a weighted mean and weighted covariance matrix as in equations \eqref{weightedestimators} and \eqref{CovWeightedMCD}. The outlyingness can then be calculated using formula (\ref{outlyingness}).

If it is still flagged as an outlier, proceed to the next value of $\eta$ and re-estimate the SNIPLS model on the original data set, determine which variables contribute to outlyingness, and check if the observation is still an outlier. Repeat this procedure until the case is no longer outlying. The procedure, called SPADIMO, is outlined in Algorithm 1. \\

\fbox{\begin{minipage}{0.9\textwidth}
{\bf Algorithm 1: SPADIMO (Sparse Direction of Maximal Outlyingness Estimation)}\\
Let $\boldsymbol{X}$ denote the data matrix (dimension $n \times p$) and $\boldsymbol{w}$ be a vector of case weights obtained from a given robust outlier detection procedure. Let $\mathcal{L} = [\ell_1,\ell_2]$ be a grid of values within $[0,1]$.\\ 
From these data, first standardize $\Xb$ to $\Zb$ by subtracting a robust estimate for location (e.g. weighted mean or columnwise median) and dividing by robust scale estimate (e.g. columnwise $Q_n$ scale estimator of \citet{rousseeuw1993alternatives}). If the weight $w_i$ of the observation to which we want to apply our method, is equal to zero, then replace that weight by a very small weight (e.g. 0.0001). Then construct $\Zb_w = (\sqrt{w_1}\zb_1^T, \ldots , \sqrt{w_n}\zb_n^T)^T$ and $\yb_w$ as outlined in Section \ref{sec:outregprob}. Set $\Zb^{(\eta)}=\Zb_w$ to start the algorithm.
Decreasing from $\ell_2$ to $\ell_1$, for each $\eta \in \mathcal{L}$:
\begin{itemize}
\item Obtain $\boldsymbol{b}_{\eta}$, the sparse PLS vector of regression coefficients regressing $\Zb^{(\eta)}$ on $\boldsymbol{y}_w$ at $h = 1$.
\item Determine $v = \{j: \boldsymbol{b}_{\eta, j} \neq 0\}$, the subset of variable(s) contributing to outlyingness.
\item Update $\Zb^{(\eta)} = \Zb^{(\eta)} \setminus \{Z_v\}$, with $Z_j$ denoting the $j{\mathrm{th}}$ column of $\boldsymbol{Z}$.
\item Compute $o(\zb^{(\eta)}_i)$, where $\zb_i^{(\eta)}$ denotes the $i{\mathrm{th}}$ row of $\Zb^{(\eta)}$ 
\end{itemize}
Stop the algorithm if $o(\zb_i^{(\eta)})^2 < \chi^2_{\alpha, q}$, where $\alpha$ denotes the required $\chi^2$ significance level and $q$ denotes the number of remaining columns of $\Zb^{(\eta)}$.
\end{minipage}}
\vspace{5mm}
 
The algorithm is sensitive to the initial choice of $\mathcal{L}$. Reasonable values for $\mathcal{L}$ will be provided from the simulation study. It is strongly advised not to scan the entire range [.01,.99], which may cause the algorithm to break off either too early or too fast. Besides, note that the approach suggested in the SPADIMO algorithm could be applied similarly using another sparse regression estimate such as the LASSO, at the cost of increased computational complexity.  

\subsection{Graphical tools\label{sec:graphics}}

Since the SNIPLS algorithm is computationally very efficient, the sparse PLS vector of regression coefficients can easily be obtained for a whole grid of values for $\eta$. The optimal $\eta$ can then be selected by analyzing figures which show the number of flagged variables for each grid value, and studying how the sparsity of the direction of maximal outlyingness changes depending on the sparsity parameter. A simple example is presented to illustrate these graphical tools.

\begin{figure}[!ht]
	\vspace*{-0.4cm}
	\centering
	\includegraphics[width=9cm]{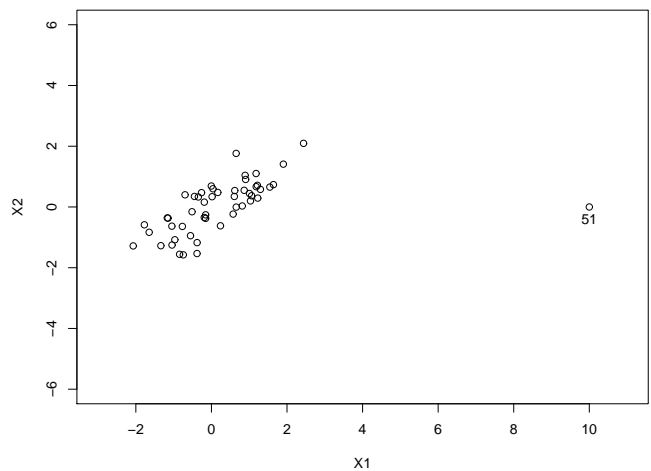}
	\caption{Data set with one outlier and $28$ noise variables.}
	\label{fig:toyexample_data}
\end{figure}

Figure \ref{fig:toyexample_data} considers $50$ points generated from a bivariate normal distribution with correlated standard normal components. One outlier ($51$) is put at position $(10,0)$. By construction, the first variable contributes most to the large outlyingness of case $51$. Next $28$ independent standard normal noise variables are added to this data set. By construction the first variable is still the only variable for which case $51$ is outlying.

\begin{figure}[!ht]
	\centering
	\vspace*{-0.4cm}
	\includegraphics[width=12cm]{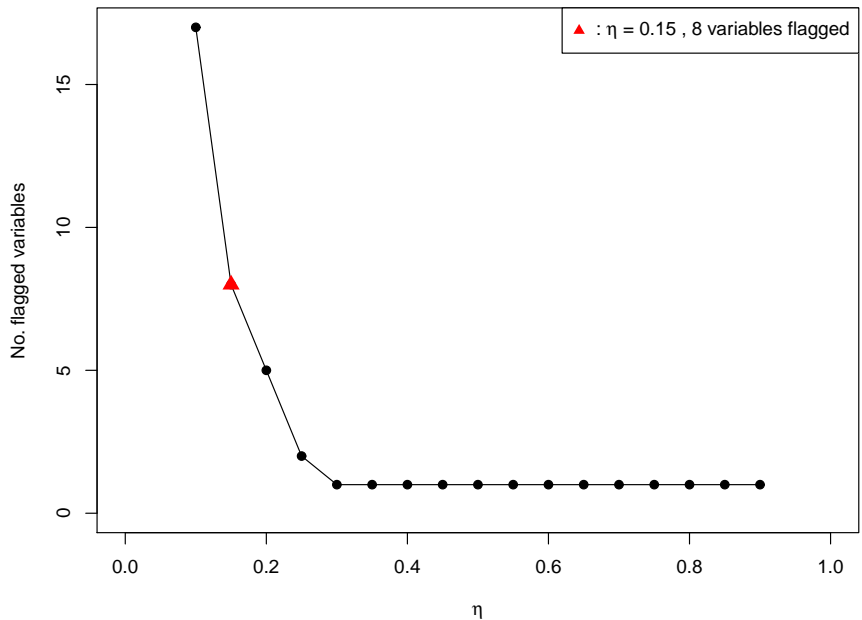}
	\caption{Number of flagged variables versus sparsity parameter. The result of the automatic approach to determine an optimal $\eta$, as described in Section \ref{sec:optimal}, is indicated by the red triangle ($\eta=0.15$ and $8$ variables are flagged).}
	\label{fig:toyexample_screeplot}
\end{figure}

The MCD estimator is used to obtain weights for each observation and then SPADIMO is applied on case $51$ with $\eta$ belonging to the grid $\{0.1,0.15,\ldots,0.9\}$. For small values of $\eta$ the direction of maximal outlyingness becomes less sparse, whereas it contains more zero components when $\eta$ is close to $1$. Figure \ref{fig:toyexample_screeplot} shows the number of variables that are flagged as outlying by SPADIMO for different values of $\eta$ and is akin to a screeplot. For $\eta\in\{0.3,\ldots,0.9\}$, SPADIMO identifies one outlying variable while several variables are flagged for $\eta <0.3$. The screeplot can be used to select the number of variables that contribute most to the outlyingness of the outlier. This can be achieved by identifying an interval for $\eta$ for which the number of flagged variables remains more or less constant. Based on Figure \ref{fig:toyexample_screeplot}, we would indeed select only one variable.

\begin{figure}[!ht]
	\centering
	\vspace*{-1.0cm}
	\hspace*{-0.1cm}
	\includegraphics[width=16cm]{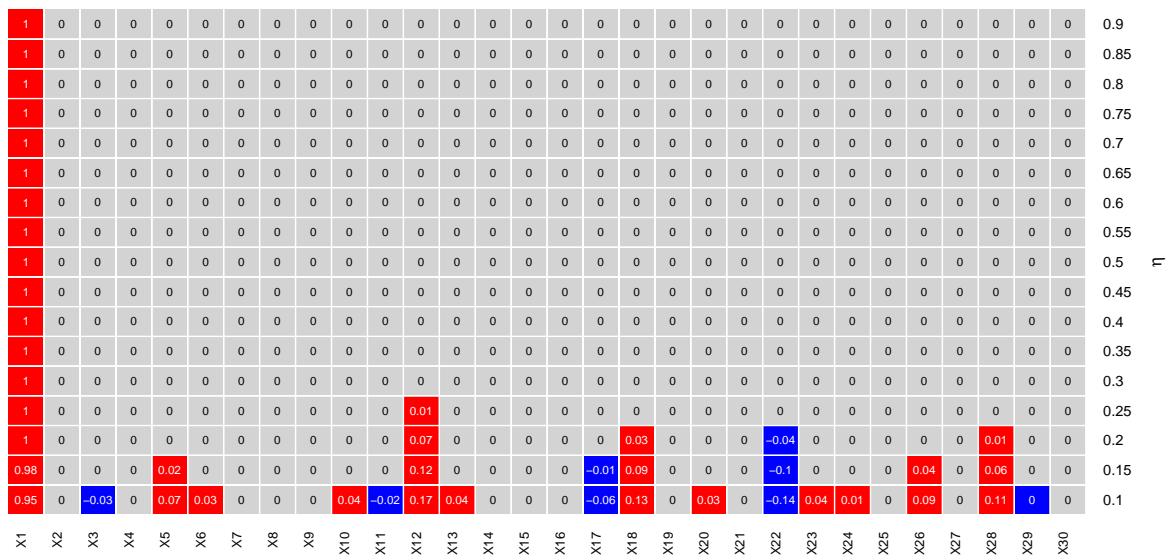}
	\vspace*{-1.5cm}
	\caption{Estimated sparse direction of maximal outlyingness for different values of $\eta$.}
	\label{fig:toyexample_direction}
\end{figure}

Figure \ref{fig:toyexample_direction} shows how the sparse direction of maximal outlyingness changes depending on $\eta$. The flagged variables correspond with the nonzero components of this direction. Figure \ref{fig:toyexample_direction} is a heatmap wherein positive components (i.e. SPLS regression coefficients) are coloured red and negative components are in blue. This reflects whether the variable is respectively outlying upwards or downwards. The first variable is clearly identified as causing the outlyingness of observation $51$ since it is the only nonzero component for $\eta\in\{0.3,\ldots,0.9\}$. For very small values of $\eta$, e.g. $0.15$, we see that there are other nonzero components, however they are very small (in absolute value) compared to the first one and thus their contribution to the outlyingness is clearly negligible. Both the screeplot and the heatmap are graphical tools to enhance the interpretation when analyzing outliers.

\section{Simulation Study \label{sec:simstudy}}
In this Section, the performance of SPADIMO using artificial data, is investigated. Since the results of various settings considered led to similar results and conclusions, only a part of our extensive simulation study is reported here. 

\paragraph{Data generation setup.} In the simulation experiment, $1000$ data sets of size $n$ were generated from a $p$-variate gaussian distribution, with $n\times p$ taken as either $500\times 50$, $200\times 200$, $50\times 500$ or $50\times 5000$.

The correlation matrices were generated randomly following \cite{agostinelli2015robust}, henceforth ALYZ, to ensure that the performance is not tied to a particular choice of correlation matrix. The ALYZ random correlation matrices yield relatively low correlations between the variables and therefore \cite{rousseeuw2017detecting} proposed the A09 correlation matrices, given by $\rho_{jh}=(-0.9)^{|h-j|}$. These matrices yield both high and low correlations. 

To contaminate the data sets, one observation is consecutively replaced by an outlier. In order to test the efficiency of the SPADIMO algorithm at detecting individual variables that contribute to the outlier's outlyingness, it is of course imperative that the outliers generated are only outlying along a subset of the variables. In order to create outliers along a few of their variables, the strategy from \cite{rousseeuw2017detecting} is adopted and we randomly replace $\lceil \varepsilon p \rceil$ of its variables by a value $\gamma$, which was varied to study its effect. In this study, the fraction of $\varepsilon$ is set to either $5\%, 10\%$ or $25\%$.

\paragraph{Evaluation setup.} For each generated data set, SPADIMO was run on the contaminated observation with $\eta$ starting at $0.9$ when $n>>p$ and starting at $0.6$ otherwise. These starting values are derived from the extended simulation study. Using these values, the number of variables that are flagged by SPADIMO were always very close to the real number of cells that have been contaminated.

In a simulation study one needs performance measures in order to evaluate the performance of the method. Therefore, the measures itemized below are calculated and the average values over 1000 simulation runs are reported in Tables \ref{table:A09} and \ref{table:ALYZ}. Note that all results reported correspond to SNIPLS estimates using a single component, since using more components did not improve the results significantly (see the Appendix \ref{sec:AppB} for the simulation results obtained using two components instead of a single one).  

\begin{itemize}
\item \# flagged: How many variables are flagged as outlying.
\item detected: How many of the outlying variables are detected (in $\%$; optimal value is 100).
\item swamped: How many good variables are flagged as outlier (in $\%$, optimal value is 0).
\item $\eta$: optimal value of $\eta$ that is used.
\end{itemize}

\begin{table}[!ht]
\scriptsize
	\centering
	\begin{tabular}{l r r r r r r r r r r r}
		\hline
		&  \multicolumn{3}{ c }{$5\%$} & & \multicolumn{3}{ c }{$10\%$} & & \multicolumn{3}{ c }{$25\%$}  \\
		&  \multicolumn{1}{ c }{$\gamma=3$} & \multicolumn{1}{ c }{$\gamma=4$} & \multicolumn{1}{ c }{$\gamma=5$} & & \multicolumn{1}{ c }{$\gamma=3$} & \multicolumn{1}{ c }{$\gamma=4$} & \multicolumn{1}{ c }{$\gamma=5$} & & \multicolumn{1}{ c }{$\gamma=3$} & \multicolumn{1}{ c }{$\gamma=4$} & \multicolumn{1}{ c }{$\gamma=5$}  \\ \cline{2-4}\cline{6-8}\cline{10-12}
		$500\times50$, A09 &  &  &  &  &  &  &  &  &  &  &    \\
		\multicolumn{1}{ c }{\# flagged}  &  3.585  &  3.381  &  3.402  & & 5.558 &   5.367 & 5.377 &   & 13.455 &  13.289 & 13.292   \\
		\multicolumn{1}{ c }{\textbf{ }detected (\%)} & 99.800 & 100.000 &  100.000  &  &   99.760  &  100.000&  100.000 &  & 99.808  &   100.000 & 	100.000   \\
		\multicolumn{1}{ c }{swamped (\%)} & 1.257   & 0.811  &  0.855  & & 1.267 &   0.816 & 0.838 &  &  1.297  &  0.781 & 0.789  \\ 
		\multicolumn{1}{ c }{$\eta$} &  0.872 &  0.863 &  0.855 &   &  0.864 &  0.856  & 0.850&  &   0.842 &  0.839 & 	0.835   \\ \hline
		$200\times200$, A09 &  &  &  &  &  &  &  &  &  &  &    \\
		\multicolumn{1}{ c }{\# flagged} &  19.177  &   12.468 & 11.309 &   &  27.959 &  22.154 & 21.145 &  &    56.019 &  51.601 & 50.894   \\
		\multicolumn{1}{ c }{\textbf{ }detected (\%)} & 99.950 &   100.000 & 100.000 &  & 99.990  &   100.000 & 100.000 &  & 100.000 & 100.000 & 100.000   \\
		\multicolumn{1}{ c }{swamped (\%)} &   4.833 &  1.299 & 0.689 &  &  4.423 &  1.197 & 0.636 &  &   4.013 & 1.067 & 0.596  \\ 
		\multicolumn{1}{ c }{$\eta$} &   0.599 &   0.592 & 	0.582 &  &   0.599 &   0.591 & 	0.581 &  &   0.598 &   0.590 & 	0.581   \\ \hline 
		$50\times500$, A09 &  &  &  &  &  &  &  &  &  &  &    \\
		\multicolumn{1}{ c }{\# flagged} &  38.195 &  32.118 &	31.866 &  &   59.717 &  55.920 & 56.284 &  &  130.329 &  129.172 &	130.129  \\
		\multicolumn{1}{ c }{\textbf{ }detected (\%)} &  94.964 & 97.360 & 98.140   &  &   93.826 &  97.208 & 98.176 &   &    95.230 &   98.131 & 99.006   \\
		\multicolumn{1}{ c }{swamped (\%)} &  3.043  &     1.637 & 	1.543 &  &  2.845 &   1.626 &1.599 &   &  3.011 &   1.735 & 1.699 \\ 
		\multicolumn{1}{ c }{$\eta$} &   0.584 &  0.552 & 0.521 & &   0.574 &   0.538 &	0.506 &  &  0.543&     0.506  &		0.477    \\ \hline
		$50\times5000$, A09 &   & & &   &  &  &  &  &  &  &    \\
		\multicolumn{1}{ c }{\# flagged} &  265.507 &  254.148 & 254.747 &  &   503.546 &  504.376 &504.480 &  &  1275.086 &  1258.361 & 1256.832   \\
		\multicolumn{1}{ c }{\textbf{ }detected (\%)} &    80.701 &  92.761 & 	95.890 &  &  86.320 & 95.810 &	97.999 &  &   94.981&    98.594 & 99.360  \\
		\multicolumn{1}{ c }{swamped (\%)} &  1.342  &    0.468 &	0.316 &   &  1.599 &  0.563 & 	0.322 &  &  5.019 &     1.406 & 0.640   \\ 
		\multicolumn{1}{ c }{$\eta$} &    0.576 & 0.544 & 	0.516 &  &    0.544 &  0.508 & 	0.482 &  &   0.488  &     0.462 &	0.438 \\ \hline
	\end{tabular}
	\caption{Results with A09 data.}
	\label{table:A09}
\end{table}
\vspace*{-0.3cm}

\begin{table}[!ht]
\scriptsize
	\centering
	\begin{tabular}{l r r r r r r r r r r r}
		\hline
		&  \multicolumn{3}{ c }{$5\%$} & & \multicolumn{3}{ c }{$10\%$} & & \multicolumn{3}{ c }{$25\%$}  \\
		&  \multicolumn{1}{ c }{$\gamma=3$} & \multicolumn{1}{ c }{$\gamma=4$} & \multicolumn{1}{ c }{$\gamma=5$} & & \multicolumn{1}{ c }{$\gamma=3$} & \multicolumn{1}{ c }{$\gamma=4$} & \multicolumn{1}{ c }{$\gamma=5$} & & \multicolumn{1}{ c }{$\gamma=3$} & \multicolumn{1}{ c }{$\gamma=4$} & \multicolumn{1}{ c }{$\gamma=5$}  \\ \cline{2-4}\cline{6-8}\cline{10-12}
		$500\times50$, ALYZ &  &  &  &  &  &  &  &  &  &  &    \\
		\multicolumn{1}{ c }{\# flagged} &    3.389  &     3.238 & 	3.250 &  &     5.296 &    5.194 & 	5.215 &  &   13.023 &   13.067 &13.135   \\
		\multicolumn{1}{ c }{\textbf{ }detected (\%)} &   97.400 &  99.567 & 99.967 &  & 97.240 &  99.400 &	99.920 &  &   97.685 &  99.385 &99.908   \\
		\multicolumn{1}{ c }{swamped (\%)} &    0.994 &  0.534 & 	0.534 &  &    0.964 &     0.498 & 0.487 & &   0.876 &  0.397 & 	0.397 \\ 
		\multicolumn{1}{ c }{$\eta$} &     0.879  &    0.868 & 	0.858 &  &     0.873 &  0.862 & 0.852 & &    0.856 &   0.846 & 	0.837  \\ \hline
		$200\times200$, ALYZ &  &  &  &  &  &  &  &  &  &  &    \\
		\multicolumn{1}{ c }{\# flagged} &   18.589 &   11.940 & 	10.610 &  &   27.452 &    21.586 & 	20.515 &  &   55.714 &  51.256 & 50.52  \\
		\multicolumn{1}{ c }{\textbf{ }detected (\%)} &  99.970 &  100.000 & 100.000 &   &   99.980 & 100.000 & 100.000 &  &   99.984 &  100.000 & 	100.000  \\
		\multicolumn{1}{ c }{swamped (\%)} &    4.522 &   1.021 &	0.321 &  &   4.142 &  0.881 & 0.286 &  &     3.815  &     0.837 & 	0.347   \\ 
		\multicolumn{1}{ c }{$\eta$} &   0.600 &  0.598 & 	0.589 &  &   0.600 &  0.598 & 0.588 &  &   0.600  &    0.595 & 	0.584  \\ \hline 
		$50\times500$, ALYZ &  &  &  &  &  &  &  &  &  &  &    \\
		\multicolumn{1}{ c }{\# flagged} &   34.858 & 27.442 &	26.260 &  &  55.970 &    51.432 & 	51.051 &  &  126.920 &  126.042 & 125.941   \\
		\multicolumn{1}{ c }{\textbf{ }detected (\%)} &  93.880  &   97.668 &	98.756 & &  92.794 &    97.316 & 98.534 &  &    94.658 &  98.523 & 	99.328  \\
		\multicolumn{1}{ c }{swamped (\%)} &    2.397  &    0.637 & 0.331 &  &   2.127 &  0.616 &0.396 & &    2.293 &    0.770 & 0.475   \\ 
		\multicolumn{1}{ c }{$\eta$} &   0.594  &    0.578 & 0.556 &   &  0.586 &    0.564 &0.537 &  &    0.552 &   0.519 & 0.492  \\ \hline
		$50\times5000$, ALYZ &  &  &  &  &  &  &  &  &  &  &    \\
		\multicolumn{1}{ c }{\# flagged} &   256.918 &  248.166 &247.305 &  &  503.867 & 501.135 &	498.420 &  &  1273.590 &  1256.568 &1250.408   \\
		\multicolumn{1}{ c }{\textbf{ }detected (\%)} &   79.974 &  93.925 & 97.304 &  &    87.204 &  96.915 & 	98.749 &  &   95.161 &  98.905 & 99.600 \\
		\multicolumn{1}{ c }{swamped (\%)} &  1.200 &   0.281 & 0.085 &  &   1.508 &   0.368 & 	0.104 &   &    2.242 & 0.540 & 	0.144   \\ 
		\multicolumn{1}{ c }{$\eta$} &   0.575 &   0.549 & 	0.524 &  &   0.541 &  0.511 & 	0.489 &  &   0.486 &  0.463 & 	0.446   \\ \hline
	\end{tabular}
	\caption{Results with ALYZ data.}
	\label{table:ALYZ}
\end{table}

\paragraph{Discussion of results.}
It can be seen that the average detection rate is always very close to $100\%$, which implies that SPADIMO is able to detect almost all contaminated variables. This does not come at the expense of wrongly flagging clean variables, since the swamping rate remains low, illustrating the very good performance of the proposed methodology and the automatic SPLS parameter selection. The results are slightly less overwhelming when the number of variables far exceeds the number of cases ($p>>n$), but even for data of dimension $50\times5000$, still at least $80\%$ of the outlying cells are detected, while the swamping rate remains well below $5\%$. Note that there is of course a trade-off between these performances and the value of $\eta$. Depending on the application, it might be more important to detect at least all the variables contributing to the outlyingness (i.e. small $\eta$), while accepting a few too many; or perhaps one only wants to flag the most important variables and certainly not too much (i.e. large $\eta$). The graphical tools described in Section \ref{sec:graphics} may certainly be helpful to make a decision. This effect can also be tuned by using different starting values for $\eta$. Deriving from the extended simulation study that led up to the results reported here, we recommend to start at $0.9$ when $n>>p$ and at $0.6$ otherwise. Note that the optimal value of $\eta$ is very close to the suggested starting value. Furthermore, the performance is similar when using ALYZ or A09 for generating the correlations, which indicates that the methodology works for both highly and moderately correlated data. For a data set of dimension $50\times5000$, it takes less than 3 seconds to run SPADIMO for a single case, while it takes around a second or even less for data sets with few dimensions. The required resources were measured on an Intel Core i5 with 2.7 GHz and 8 GB RAM.

\section{Examples \label{sec:example}}

\subsection{Top Gear Data}
The data for the first example were taken from the website of the popular British television show Top Gear by \cite{alfons2012robusthd}. It consists of 297 cars quantified in 32 variables, but as in \cite{rousseeuw2017detecting}, we will only focus on the 11 objectively measured numerical variables. Five of these variables (such as top speed) were logarithmically transformed first, since they were skewed. Moreover, 52 cars with missing values are omitted, resulting in a data set with 245 observations and 11 variables.
 
\begin{figure}[!ht]
	\centering
	\hspace*{-1.2cm} 
	\includegraphics[width=18cm]{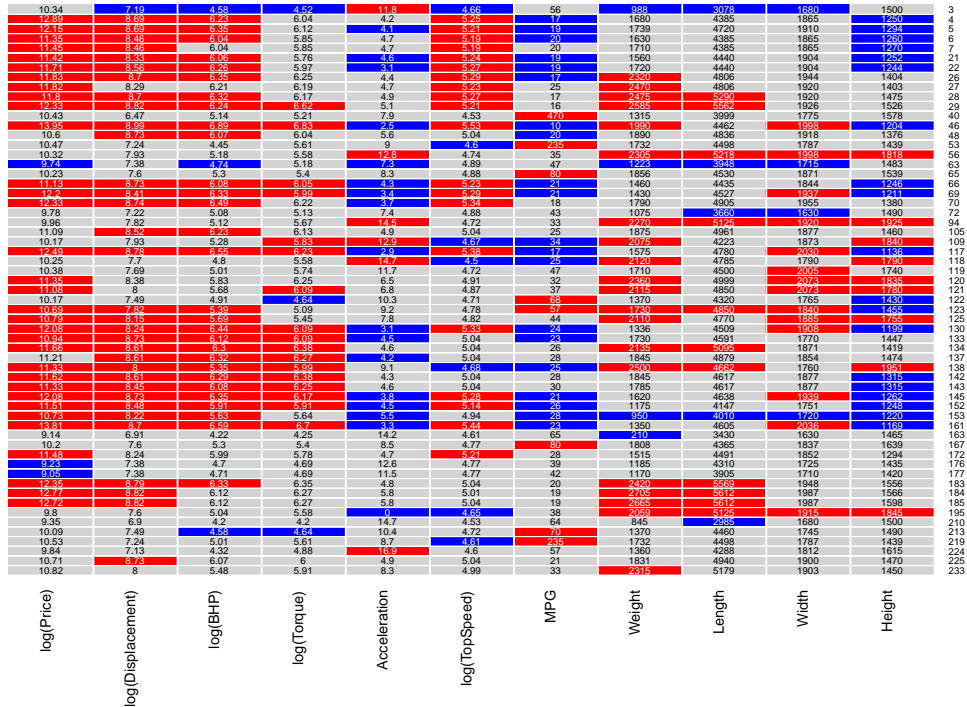}
	\caption{Heatmap of outliers (detected by MCD) in the Top Gear data set. The red and blue boxes indicate the variables that are flagged as outlying variables by SPADIMO.}
	\label{fig:heatmaptopgear_full}
\end{figure}

Since $n>p$, the MCD estimator is applied as a robust estimator for location and scale and the case weights are calculated as in formula \eqref{weights}. As a result, 59 out of 245 observations are flagged as outliers and these will be studied using SPADIMO so as to detect which variables contribute most to their outlyingness. 
The results are plotted as a heatmap in Figure \ref{fig:heatmaptopgear_full}, where the 59 outliers are represented as rows. 
For every outlier detected by MCD, the individual cells and corresponding values are shown. The individual cells that contribute most to the outlyingness, according to the SPADIMO algorithm, are shown as a colored box. The anomalous variables whose corresponding component in the sparse direction of maximal outlyingness is positive are colored red and those with a negative component are in blue. It can immediately be seen that some outliers are deviating in a lot of cells, whereas others only have an atypical value for a few cells. 

\begin{figure}[!ht]
	\centering
	\hspace*{-0.8cm} 
	\includegraphics[width=18cm]{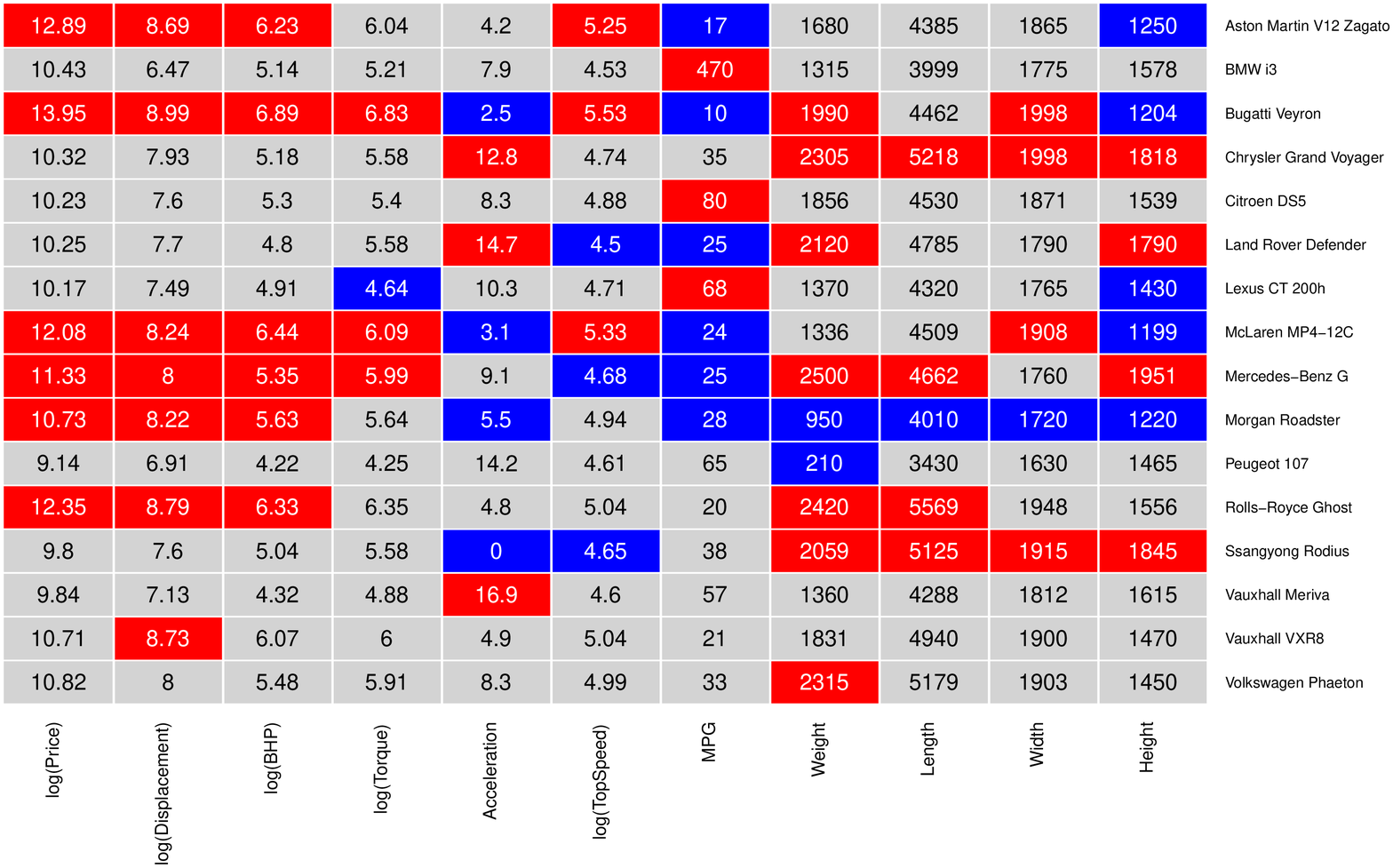}
	\caption{Heatmap of some outliers (detected by MCD) in the Top Gear data set. The red and blue boxes indicate the variables that are flagged as outlying variables by SPADIMO.}
	\label{fig:heatmaptopgear_selected}
\end{figure}

Let us focus on some examples (see Figure \ref{fig:heatmaptopgear_selected}). 
From this analysis, it can be seen that some outliers only have atypical values in a single column, such as the BMW i3 (MPG of 470), Citro\"{e}n DS5 (MPG of 80), Lexus CT 200h (log torque of 4.64), Peugeot 107 (weight of 210), Vauxhall Meriva (acceleration of 16.9), Vauxhall VXR8 (log displacement of 8.73) and Volkswagen Phaeton (weight of 2315). The weight of the Peugeot 107 is clearly an error, but not all of these atypical values are errors, since for example the BMWI i3 is an electrical vehicle with a small additional gas engine which explains its very high MPG.

Moreover, a fair amount of cars are multivariate outliers. None of their properties are unusual in the univariate sense, but in combination with other characteristics, the corresponding cars are flagged as outliers.
The SPADIMO analysis can definitely distinguish between both types of outliers; moreover the color of the anomalous variables reflects whether the observed value is outlying upwards or downwards, which helps interpreting this complexly structured data set. 

\begin{figure}[!ht]
	\centering
	\vspace*{-0.2cm} 
	\includegraphics[width=12cm]{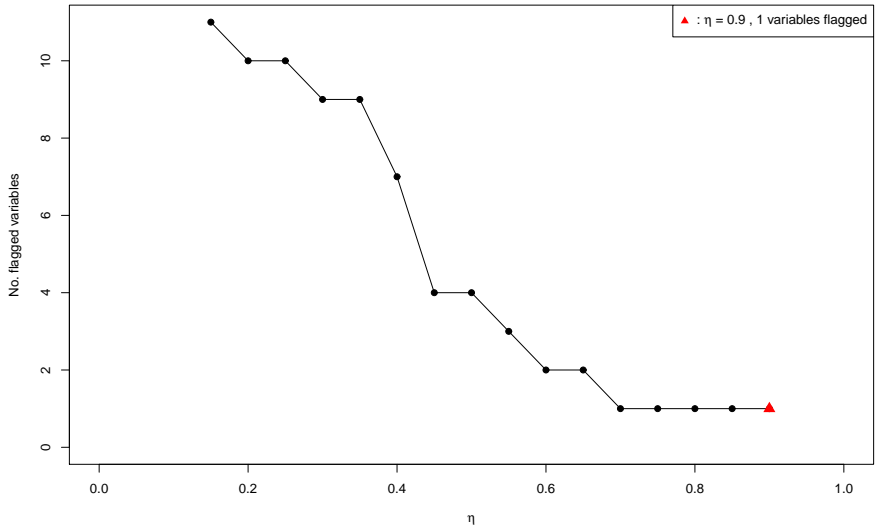}
	\caption{Number of flagged variables versus sparsity parameter for the Peugeot 107.}
	\label{fig:Peugeot107_screeplot}
\end{figure}
\begin{figure}[!ht]
	\centering
	\vspace*{-1.2cm}
	\includegraphics[width=15cm]{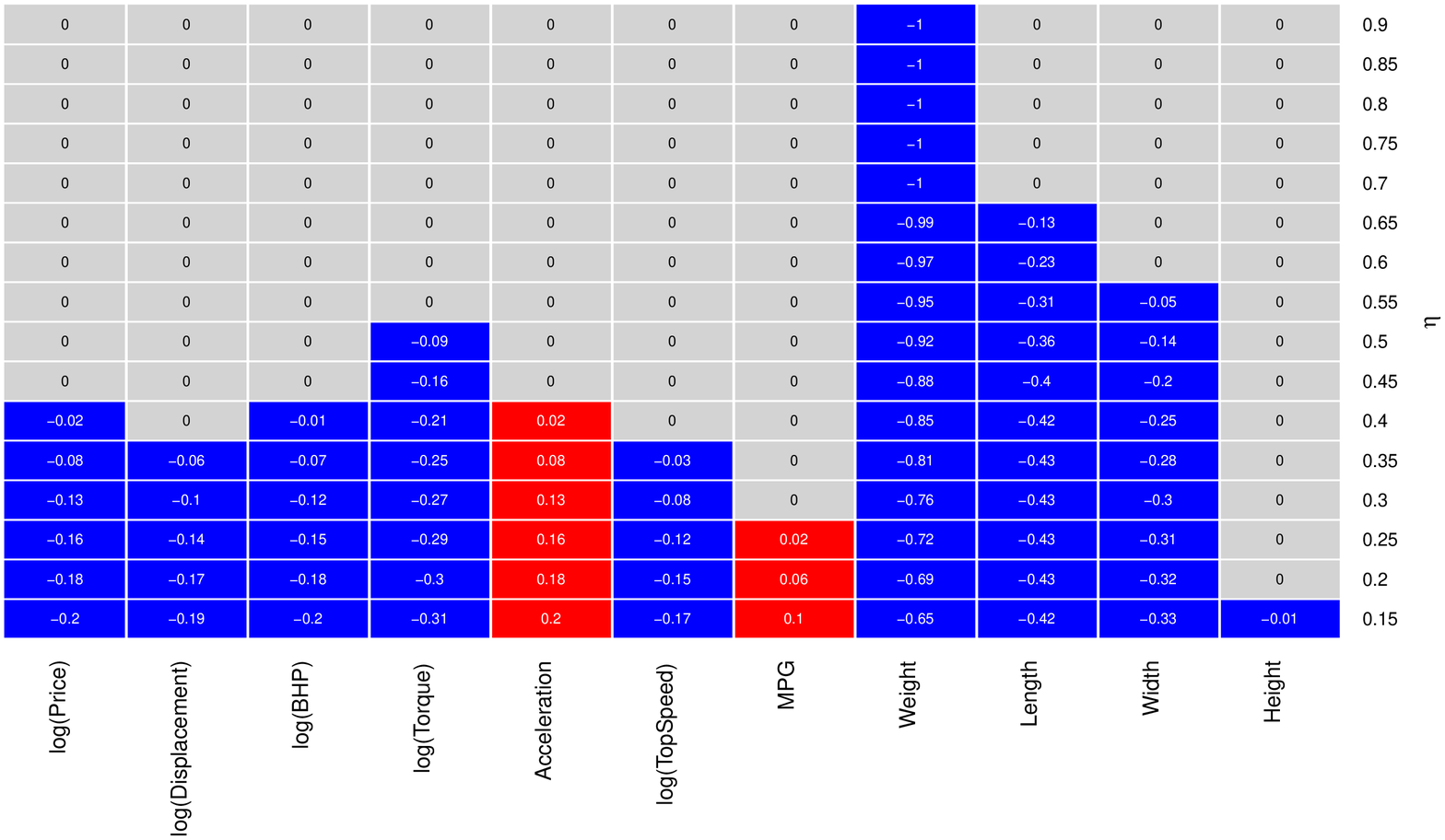}
	\caption{The estimated sparse direction of maximal outlyingness of the Peugeot 107 for different values of $\eta$.}
	\label{fig:Peugeot107_direction}
\end{figure}
\begin{figure}[!ht]
	\centering
	\vspace*{-0.4cm}
	\includegraphics[width=15cm]{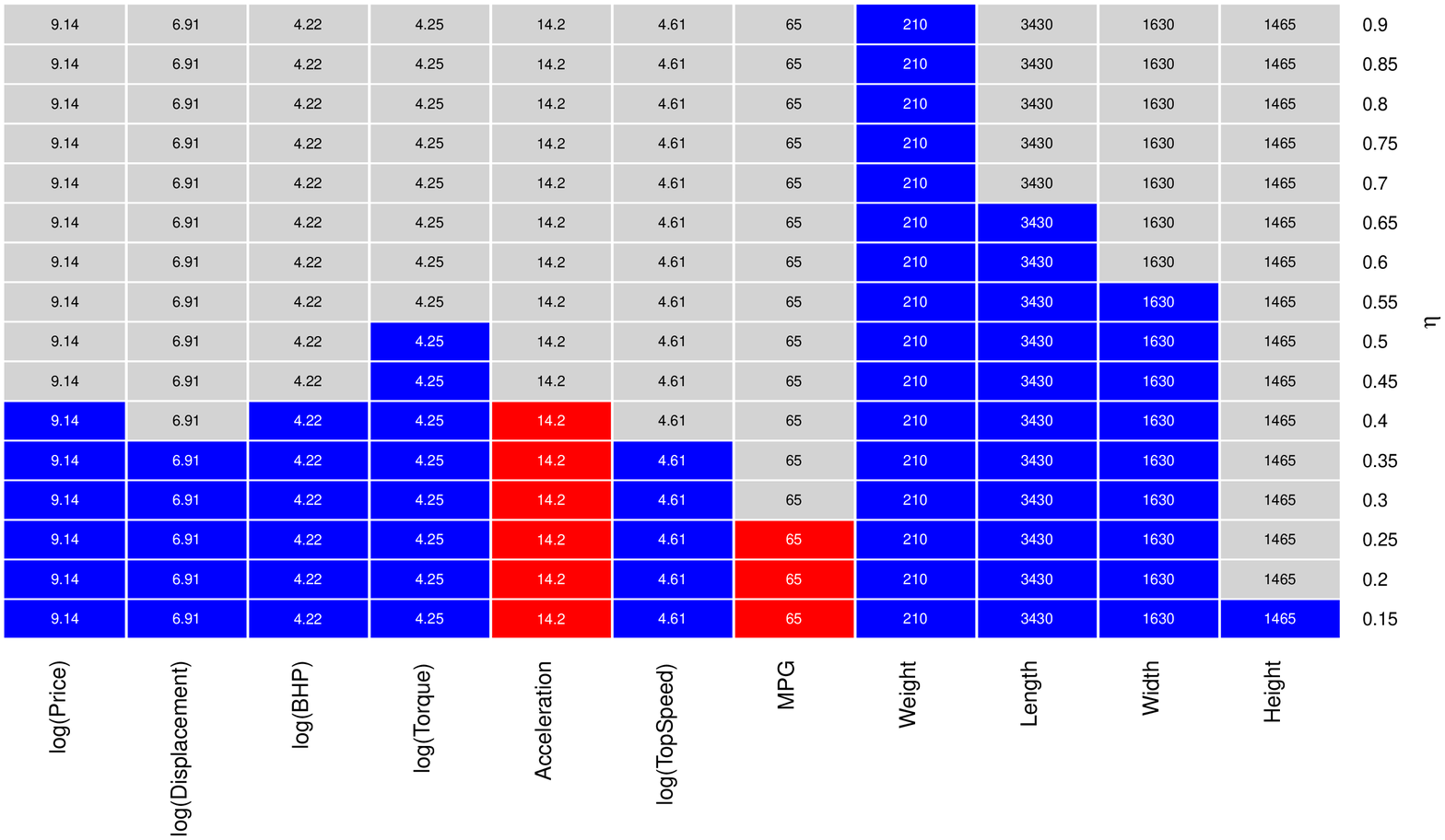}
	\caption{The anomalous cells of the Peugeot 107 as flagged by SPADIMO for varying $\eta$.}
	\label{fig:Peugeot107_case}
\end{figure}

Let us have a closer look at the Peugeot 107. Figure \ref{fig:Peugeot107_screeplot} shows the number of flagged variables for varying values of $\eta$. For $\eta>0.4$, the number of identified variables ranges from $1$ to $4$. The method itself, as described in Algorithm 1, selects a single variable at $\eta=0.9$. Figure \ref{fig:Peugeot107_direction} indicates that variable \textit{weight} causes mostly the outlyingness of Peugeot 107 as it is the first variable to be flagged. Other variables that seem to contribute to the oulyingness are \textit{length}, \textit{width} and \textit{log(torque)}. The Peugeot 107 is a small city car as can be seen from the values listed in Figure \ref{fig:Peugeot107_case}. More variables are flagged when $\eta<0.4$, but their corresponding components remain rather small (in absolute value) so they clearly contribute less to the outlyingness of this car.

\subsection{Glass Data\label{sec:glass}}

The second example concerns a data set consisting of 180 electron probe X-ray microanalysis (EPXMA) spectra of arch\ae ological glass vessels excavated in the Anwerp, Belgium area, further referred to as the {\em glass data}. The data set has been described extensively before. At first, detailed information on the analytical chemistry involved, as well as quantification based on software involving manual peak selection and integration, can be found in \cite{Janssens}. Secondly, \cite{Lemberge} have illustrated that it is possible to skip the tedious manual peak selection and quantification step by multivariate calibration. They applied partial least squares regression with very promising results, regarding each of the major chemical constituents targeted. However, in \cite{Serneels:PRM} it is mentioned that the results in \cite{Lemberge} were only obtained after a set of outliers had been removed from the data, that correspond to samples measured with a different detector efficiency. In \cite{Serneels:PRM}, it is then shown that the analytical procedure could have been sped up further by applying partial robust M-regression, a therein newly proposed robust alternative to partial least squares regression, instead of the the manual outlier removal process. 

Details on each of these analytical steps can be retrieved in the corresponding papers. However, in order to illustrate the effectiveness of the presently proposed method, there are a few key facts to keep in mind about the glass data. The data consist of 180 EPXMA spectra measured at 750 energy channels, out of which all cases beyond case 143 are outliers measured with a different detector efficiency. On top of that, the data have been reported to consist of four clusters corresponding to four glass types: {\em sodic}, {\em potassic}, {\em calcic} and {\em potasso--calcic} glass \citep{Janssens}. The vast majority of the data correspond to sodic glass vessels. 

\begin{figure}[!ht]
\centering
\includegraphics[width=.6\textwidth,height=.7\textheight]{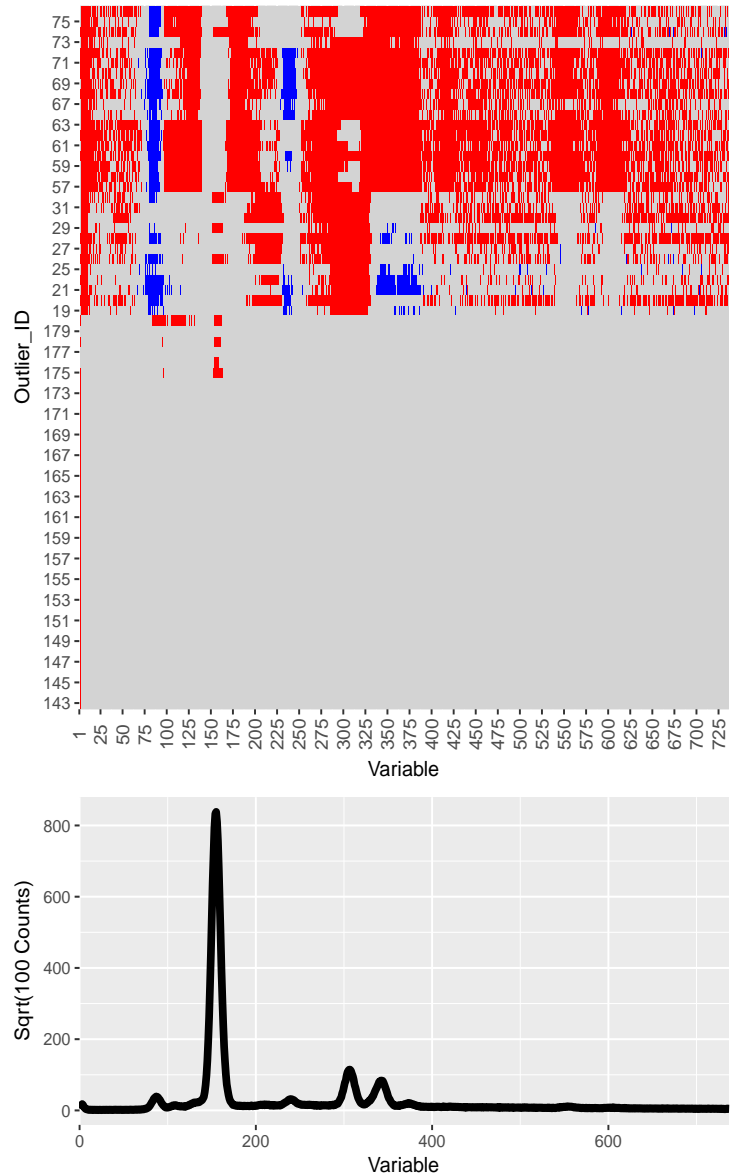}
\caption{Top: Heatmap of outlying cells in outliers in the glass data set. The black boxes indicate the variables that are flagged as outlying variables by SPADIMO (single latent variable). Bottom: For illustrative purposes, EPXMA spectrum corresponding to outlier 145.}
\label{fig:heatmapglass}
\end{figure}

Outlier detection was carried out by means of ROBPCA \citep{Hubert:ROBPCA} which detects 68 outliers. This set of outliers contains all 38 measurement outliers, but it also contains the cases corresponding to the non-sodic glass vessels. In the top of Figure \ref{fig:heatmapglass}, a heatmap is plotted showing the individual cells detected as contributing to outlyingness by SPADIMO in the cases detected as outliers by ROBPCA. The parameters used in the SPADIMO scan are: $\mathcal{L}=[.1,.6]$ and $\alpha=.99$. 

The cells that correspond to variables that contribute most to outlyingness, are plotted in black. In fact, the result from SPADIMO is interpretable, knowing the nature in which the outliers deviate. At first,  Figure \ref{fig:heatmapglass} shows a clear difference between the top group of outliers, that have outlying cells more or less across the entire range of energies, whereas the bottom group of outliers contain outlying cells only in very specific areas. By analyzing the case numbers, it is clear that the former correspond to the non-sodic glass vessels, whereas the latter ones correspond to the measurement outliers. The non-sodic glass vessels have a different chemical composition, not only characterized by the main characteristic constituents such as sodium and calcium, but also by having a different matrix, containing different trace elements. Hence it is very plausible that deviations may be present across the entire spectrum when compared to sodic glass. Moreover, it is even possible to detect the individual types of glass the outliers correspond to based on the SPADIMO analysis. When looking at the cells that correspond to variables roughly in the range 300 through 375, one can see that the non-measurement outliers (case number $< 143$) can be subdivided into three groups: for some cases, only roughly variables 300 through 330 are outlying (left hand black block), whereas for another group only roughly variables 330 through 375 are outlying (right hand black block). For some cases, the entire range 300--375 is detected as outlying. This phenomenon has a straightforward interpretation: the variable range 300--330 corresponds to the K $K_\alpha$ peak, whereas the range 330-375 corresponds to the Ca $K_\alpha$ and $K_\beta$ peaks. Cases where the 300--330 range is outlying, correspond to the potassic samples. Cases where the 330--375 range is outlying, correspond to the calcic samples. Cases where both are outlying, correspond to the potasso--calcic samples. Note that a similar discrepancy can be observed in variable range 540--620. This range corresponds to iron and manganese peaks, elements of which the different glass types typically also contain different amounts. 

Finally, the measurement outliers can also clearly be distinguished from the compositional outliers. Detector efficiency is related to how X-rays are absorbed by the detector. The more energetic X-rays are, the more likely they will penetrate an inefficient detector. Therefore, a slight decrease in detector efficiency like the one witnessed in this data set, has no effect on high energy X-rays. As energy increases toward the right hand side of the plot, it is logical that all outlying cells are being detected at the left hand side. For all measurement outliers, the detector efficiency issue has generated an artifact at the beginning of the spectrum (very low energetic X-rays). This artifact is being detected by SPADIMO. Further down the spectrum, the fact that the effect can be detected or not, doesn't just depend on the detector efficiency, but also on the individual element concentrations. A glass vessel that already contains a naturally low end concentration of sodium, will end up being outlying in the corresponding range with an inefficient detector. However, a high sodium vessel might end up in the majority of the data with lower detector efficiency. For these reasons, one can see from Figure \ref{fig:heatmapglass}, that only for selected samples, outlying cells aren't just detected in the artifact range at the extreme left end of the spectrum, but also in a few of the peaks corresponding to low elements with low characteristic energies such as sodium and silicon.

\section{Further applications and conclusions\label{sec:conclusion}}

In this paper, it has been proven that calculation of the direction of maximal outlyingness can be rewritten as a regression problem. It has been shown that applying variable selection methodology to this regression problem leads to detection of individual variables that contribute to a case's outlyingness. Therefore, it eventually helps understanding why an outlier lies out. An algorithm, called SPADIMO, has been proposed to accomplish this topic in practice. We also present two graphical tools that are helpful to gain insight in the studied observation. SPADIMO is based on the estimation of the regression problem associated with outlyingness by sparse NIPALS regression. The latter is a multivariate regression technique for integrated variable selection and regression, that is both suitable for low and high dimensional data. By consequence, the proposed SPADIMO algorithm can be applied with equal convenience to both data configurations. An implementation of SPADIMO will be made publicly available as an \textsf{R} package. 

In an extensive simulation study, the method has been proven to by and large detect exactly those outlying cells that had been set to outlying in simulated data. Even for very high dimensional data, SPADIMO does detect over $80\%$ of cells truly contributing to simulated cellwise outliers, whereas it only yields less than $5\%$ false positives. 

SPADIMO can turn out to be of great practical importance for various fields of science. One can think of the detection of transfer fraud, where fraudulent transactions are both outliers and the most interesting cases at the same time, and where it is of utmost importance to be able to analyze in which way the transaction is fraudulent. In bioinformatics, one often wants to distinguish cancer cells from regular ones. The cancer cells will have different expressions in just a subset of the genes. Therefore, they will be outlying with respect to the bulk of the data, and it is even more interesting to know which are the outlying genes. In process chemistry, a plant often produces out-of-specification product without an {\em obvious} established cause according to experienced operators. In that case, the off-spec production steak will be outlying in a multivariate way, and SPADIMO can help select which combination of variables to tune. Even though the aforementioned examples are all very realistic, the examples shown in this article are yet of another nature. It was shown that SPADIMO allows to distinguish univariate from multivariate oultiers in the Top Gear data set, which helps to analyze the different types of cars in it more efficiently. In a second data set, SPADIMO allows to detect outlying cells in a data set of arch\ae ological glass vessels. There, the outcome from SPADIMO can perfectly be traced back to the nature of the individual outliers: some outliers belong to a deviating glass type, whereas others can be traced back to measurement error. In these two examples of different natures, it has been illustrated that SPADIMO yields highly interpretable information regarding individual outliers.

SPADIMO can be a great tool to enhance interpretation when analyzing outliers. It can even have more potential than being a standalone tool used in one-off analyses. It can, for instance, become integrated in software packages, showing variables contributing to outlyingness on a one click basis. It can become an ancillary part of cellwise robust estimation procedures, which is a path of research yet to be explored.

\appendix
\section{Appendix: Proofs \label{sec:AppA}}
\subsection{Proof of Proposition \ref{prop1}\label{sec:AppA1}}

\begin{proof}
Note that our weighted covariance matrix $\hbS_w$, like all covariance matrices, is a positive-semidefinite matrix. Since we also assume it is not singular and $\hbS_w^{-1}$ exists, we know that $\hbS_w$ is positive-definite. We now apply the Cauchy-Bunyakovskiy-Schwarz inequality to $\xb = \hbS_w^{-1/2}\xb_1$ and $\yb = \hbS_w^{1/2}\yb_1$, for arbitrary $\xb_1,\yb_1 \in \R^p$.
This results in the following inequality
\begin{equation} \nonumber
(\xb_1^T\yb_1)^2 \leq \xb_1^T\hbS_w^{-1}\xb_1 \yb_1^T \hbS_w\yb_1
\end{equation}
We have equality if $\yb = c\xb$ with $c\in\R$, which means $\hbS_w^{1/2}\yb_1 = c \hbS_w^{-1/2}\xb_1$ or $\yb_1 = c \hbS_w^{-1}\xb_1$. 
So summarized, for any $\xb,\yb \in \R^p$ we have the inequality
\begin{equation} \nonumber
(\xb^T\yb)^2 \leq \xb^T\hbS_w^{-1}\xb \yb^T \hbS_w\yb,
\end{equation}
where there is equality if and only if $\yb = c \hbS_w^{-1}\xb$.

We now look at $$\frac{(\xb^T\ab - \hbm_w^T\ab)^2}{\ab^T\hbS_w\ab}$$ and apply this inequality:
\begin{equation} \nonumber
\frac{((\xb - \hbm_w)^T\ab)^2}{\ab^T\hbS_w\ab} \leq \frac{(\xb - \hbm_w)^T \hbS_w^{-1}(\xb - \hbm_w) \ab^T \hbS_w \ab}{\ab^T\hbS_w\ab} =(\xb - \hbm_w)^T \hbS_w^{-1}(\xb - \hbm_w).
\end{equation}
We have equality in the above inequality if $\ab = c \hbS_w^{-1}(\xb - \hbm_w)$. So $$\ab = \frac{\hbS_w^{-1}(\xb - \hbm_w)}{\|\hbS_w^{-1}(\xb - \hbm_w)\|}$$ is the direction $\ab$ that maximizes $$\frac{|\xb^T\ab - \hbm_w^T\ab|}{\sqrt{\ab^T\hbS_w\ab}}$$ and for this $\ab$ we have $$\left(\frac{|\xb^T\ab - \hbm_w^T\ab|}{\sqrt{\ab^T\hbS_w\ab}}\right)^2 = (\xb - \hbm_w)^T \hbS_w^{-1}(\xb - \hbm_w) = o(\xb)^2.$$
\end{proof}

\subsection{Proof of Theorem \ref{theorem1}\label{sec:AppA2}}
\begin{proof}
	We know that, by the theory of ordinary least squares regression, $$\thetab_{\varepsilon} = (\Xb_{w,\varepsilon}^T\Xb_{w,\varepsilon})^{-1}\Xb_{w,\varepsilon}^T\yb_{w,\varepsilon}$$ and by the definition of our weighted covariance matrix, $\hbS_{w,\varepsilon} = \frac{1}{n_{w,\varepsilon}-1} \Xb_{w,\varepsilon}^T\Xb_{w,\varepsilon}$, we can write $$\thetab_{\varepsilon} = ((n_{w,\varepsilon}-1)\hbS_{w,\varepsilon})^{-1}\Xb_{w,\varepsilon}^T\yb_{w,\varepsilon}.$$ We know that $((n_{w,\varepsilon}-1)\hbS_{w,\varepsilon})^{-1} = \frac{1}{n_{w,\varepsilon}-1} \hbS_{w,\varepsilon}^{-1}$ and it is easy to see that $\Xb^T_{w,\varepsilon}\yb_{w,\varepsilon} = \sqrt{\varepsilon}(\xb - \hbm_{w,\varepsilon})$, if we look at the definitions of $\Xb_{w,\varepsilon}$ and $\yb_{w,\varepsilon}$. Thus we have that $$\thetab_{\varepsilon} = \frac{\sqrt{\varepsilon}}{n_{w,\varepsilon}-1} \hbS_{w,\varepsilon}^{-1}(\xb - \hbm_{w,\varepsilon}).$$ Since $\varepsilon$ is strictly larger than zero, we have that $$\frac{\thetab_{\varepsilon}}{\|\thetab_{\varepsilon}\|} = \frac{\hbS_{w,\varepsilon}^{-1}(\xb - \hbm_{w,\varepsilon})}{\|\hbS_{w,\varepsilon}^{-1}(\xb - \hbm_{w,\varepsilon})\|}.$$ Then we get that $$\lim_{\varepsilon\rightarrow0}\frac{\thetab_{\varepsilon}}{\|\thetab_{\varepsilon}\|}=\frac{\hbS_w^{-1}(\xb - \hbm_w)}{\|\hbS_w^{-1}(\xb - \hbm_w)\|} = \ab(\xb)$$ since $\lim_{\varepsilon\rightarrow0}n_{w,\varepsilon}=n_w$, $\lim_{\varepsilon\rightarrow0}\hbm_{w,\varepsilon}=\hbm_w$ and $\lim_{\varepsilon\rightarrow0}\hbS_{w,\varepsilon}^{-1}=\hbS_w^{-1}$.
\end{proof}

\section{Appendix: Simulation results from two component SNIPLS models\label{sec:AppB}}

\textbf{Simulation results for $h = 2$:} Simulation results are provided for SPADIMO based on SNIPLS models with two latent variables. It can be seen that increasing the number of latent variables, does not ameliorate the results.

\subsection{Results with A09 data\label{sec:AppB1}}

\begin{table}[!ht]
	\scriptsize
	\centering
	\label{table:A09_2latent}
	\begin{tabular}{l r r r r r r r r r r r}
		\hline
		&  \multicolumn{3}{ c }{$5\%$} & & \multicolumn{3}{ c }{$10\%$} & & \multicolumn{3}{ c }{$25\%$}  \\
		&  \multicolumn{1}{ c }{$\gamma=3$} & \multicolumn{1}{ c }{$\gamma=4$} & \multicolumn{1}{ c }{$\gamma=5$} & & \multicolumn{1}{ c }{$\gamma=3$} & \multicolumn{1}{ c }{$\gamma=4$} & \multicolumn{1}{ c }{$\gamma=5$} & & \multicolumn{1}{ c }{$\gamma=3$} & \multicolumn{1}{ c }{$\gamma=4$} & \multicolumn{1}{ c }{$\gamma=5$}  \\ \cline{2-4}\cline{6-8}\cline{10-12}
		$500\times50$, A09 &  &  &  &  &  &  &  &  &  &  &    \\
		\multicolumn{1}{ c }{\# flagged}  & 6.027 & 6.303 & 6.445 & & 7.986 & 8.163 & 8.344 & & 15.915 &  15.898 & 16.038  \\
		\multicolumn{1}{ c }{\textbf{ }detected (\%)} & 99.800 & 100.000 & 100.000 & & 99.800 & 100.000 & 100.000 & & 99.862 &  100.000 & 100.000   \\
		\multicolumn{1}{ c }{swamped (\%)} & 6.453 &  7.028  & 7.330 & & 6.658 & 7.029 & 7.431 & & 7.927 &  7.832 & 8.211  \\ 
		\multicolumn{1}{ c }{$\eta$} &  0.885 & 0.884 & 0.882 & & 0.878 & 0.879 & 0.878 & & 0.853 &   0.856 & 0.856 \\ \hline
		$200\times200$, A09 &  &  &  &  &  &  &  &  &  &  &    \\
		\multicolumn{1}{ c }{\# flagged} &   39.654 & 38.000 & 38.380 & & 48.773 &   47.758 & 48.579& & 76.840 & 75.937 &  76.656 \\
		\multicolumn{1}{ c }{\textbf{ }detected (\%)} & 99.990 &  100.000 & 100.000 & & 99.985 & 100.000 & 100.000 & & 99.994 &   100.000 & 100.000  \\
		\multicolumn{1}{ c }{swamped (\%)} &   15.608 & 14.737 &14.937 & & 15.987 & 15.421 &15.877 & & 17.895 & 17.291 & 17.771 \\ 
		\multicolumn{1}{ c }{$\eta$} &  0.600 & 0.594 & 0.588 & & 0.599 &  0.594 & 0.587 & & 0.598 &  0.592 & 0.585 \\ \hline 
		$50\times500$, A09 &  &  &  &  &  &  &  &  &  &  &    \\
		\multicolumn{1}{ c }{\# flagged} &  70.558 &  77.704  & 89.005 & & 94.273 &  108.220 & 121.557& & 170.887 & 183.875 & 193.760 \\
		\multicolumn{1}{ c }{\textbf{ }detected (\%)} &  95.228 & 97.656  & 98.368 & & 94.020 & 97.330 & 98.258 & & 95.810 & 98.351 & 99.036  \\
		\multicolumn{1}{ c }{swamped (\%)} & 9.842 & 11.219 & 13.561 & & 10.503 & 13.234 &16.095 & & 13.633 & 16.250 & 18.657 \\ 
		\multicolumn{1}{ c }{$\eta$} &  0.587 &  0.559  & 0.535 & & 0.577 & 0.543 &0.516 & & 0.542 & 0.509 & 0.486   \\ \hline
		$50\times5000$, A09 &   & & &   &  &  &  &  &  &  &    \\
		\multicolumn{1}{ c }{\# flagged} & 431.331 &  499.536 & 581.579 & & 729.868 & 814.465 & 908.111 & & 1575.825 & 1601.788 & 1662.530  \\
		\multicolumn{1}{ c }{\textbf{ }detected (\%)} &  81.394 & 92.835 & 95.907 & & 86.834 & 96.105 & 98.124 & & 95.437 & 98.737  & 99.427 \\
		\multicolumn{1}{ c }{swamped (\%)} &  4.797 &    5.631& 7.196 & & 6.571 &  7.421 & 9.278 & & 10.210 & 9.802 & 11.192  \\ 
		\multicolumn{1}{ c }{$\eta$} & .576 &   0.545 & 0.518 & & 0.544 &  0.508 & 0.480 & & 0.487 & 0.460 & 0.436 \\ \hline
	\end{tabular}
	\caption{Results with A09 data from two component SNIPLS models.}
\end{table}

\newpage

\subsection{Results with ALYZ data\label{sec:AppB2}}

\begin{table}[!ht]
	\scriptsize
	\centering
	\label{table:ALYZ_2latent}
	\begin{tabular}{l r r r r r r r r r r r}
		\hline
		&  \multicolumn{3}{ c }{$5\%$} & & \multicolumn{3}{ c }{$10\%$} & & \multicolumn{3}{ c }{$25\%$}  \\
		&  \multicolumn{1}{ c }{$\gamma=3$} & \multicolumn{1}{ c }{$\gamma=4$} & \multicolumn{1}{ c }{$\gamma=5$} & & \multicolumn{1}{ c }{$\gamma=3$} & \multicolumn{1}{ c }{$\gamma=4$} & \multicolumn{1}{ c }{$\gamma=5$} & & \multicolumn{1}{ c }{$\gamma=3$} & \multicolumn{1}{ c }{$\gamma=4$} & \multicolumn{1}{ c }{$\gamma=5$}  \\ \cline{2-4}\cline{6-8}\cline{10-12}
		$500\times50$, A09 &  &  &  &  &  &  &  &  &  &  &    \\
		\multicolumn{1}{ c }{\# flagged}  &  4.784 & 4.508 & 4.391 & & 6.502 & 6.291 & 6.235 & & 14.032 &  14.027 & 14.094  \\
		\multicolumn{1}{ c }{\textbf{ }detected (\%)} & 99.267 &  99.900 & 100.000 & & 98.360 & 99.640 & 99.980 & & 97.554 &  99.354 &  99.908 \\
		\multicolumn{1}{ c }{swamped (\%)} & 3.843 & 3.215 & 2.960 & & 3.520 & 2.909 & 2.747 & & 3.649 & 3.003 & 2.989 \\ 
		\multicolumn{1}{ c }{$\eta$} & 0.892 & 0.890 & 0.887 & & 0.888 & 0.886 & 0.883 & & 0.869 &  0.865 & 0.860  \\ \hline
		$200\times200$, A09 &  &  &  &  &  &  &  &  &  &  &    \\
		\multicolumn{1}{ c }{\# flagged} &  35.274 &  28.644 & 27.056 & & 42.819 &  37.441 &36.486 & & 68.341 &  64.320 & 64.155  \\
		\multicolumn{1}{ c }{\textbf{ }detected (\%)} & 99.980 & 100.000 & 100.000 & & 99.995 &  100.000 & 100.000& & 99.980 &  100.000 & 100.000  \\
		\multicolumn{1}{ c }{swamped (\%)} & 13.303 &  9.813 & 8.977 & & 12.678 &  9.689 &9.159 & & 12.234 &  9.547 & 9.437 \\ 
		\multicolumn{1}{ c }{$\eta$} &  0.600 &   0.600 & 0.599 & & 0.600 & 0.600 & 0.597 & & 0.600 &   0.598 & 0.592 \\ \hline 
		$50\times500$, A09 &  &  &  &  &  &  &  &  &  &  &    \\
		\multicolumn{1}{ c }{\# flagged} & 63.118 &  59.951 & 64.868 & & 84.389 &  87.361 &93.564 & & 159.789 & 167.546 & 175.643  \\
		\multicolumn{1}{ c }{\textbf{ }detected (\%)} & 94.224 & 97.676 & 98.616 & & 92.808 &  97.642 & 98.544& & 95.086 &  98.623 &  99.362 \\
		\multicolumn{1}{ c }{swamped (\%)} & 8.329 & 7.480 & 8.466 & & 8.441 & 8.564 &9.843 & & 10.915 & 11.805 & 13.718 \\ 
		\multicolumn{1}{ c }{$\eta$} &  0.597 & 0.584 & 0.567 & & 0.589 & 0.564 & 0.546 & & 0.552 & 0.519 &  0.493  \\ \hline
		$50\times5000$, A09 &   & & &   &  &  &  &  &  &  &    \\
		\multicolumn{1}{ c }{\# flagged} &  413.949 &  451.988 &505.737 & & 717.890 & 767.524 & 817.685 & & 1548.338 & 1566.894& 1595.568   \\
		\multicolumn{1}{ c }{\textbf{ }detected (\%)} & 80.282 &  94.140 & 97.431 & & 87.740 &  97.073 &98.834 & & 95.461 & 99.020 & 99.628 \\
		\multicolumn{1}{ c }{swamped (\%)} &  4.489 & 4.561 & 5.519 & & 6.204 & 6.270 & 7.189 & & 9.469 & 8.777 & 9.339  \\ 
		\multicolumn{1}{ c }{$\eta$} & 0.576 & 0.549 & 0.524 & & 0.541 &   0.511 & 0.488 & & 0.486 & 0.462 & 0.445 \\ \hline
	\end{tabular}
	\caption{Results with ALYZ data from two component SNIPLS models.}
\end{table}

%
%
%
%
%

\bibliographystyle{chicago}
 
\bibliography{bib-outlyingness_03}

\end{document}